\documentstyle[preprint,aps]{revtex}

\begin{document}

\title{\bf Mixing of superconducting  $d_{x^2-y^2}$ state with 
$s$-wave states for different filling and temperature}

\author{Angsula Ghosh$^1$\thanks{Corresponding author. e-mail:
angsula@if.usp.br, Fax +55 11 3177 9080. Address: Instituto de Fisica,
Universidade de S\~ao Paulo, Caixa Postal 66318, 05315-970 S\~ao Paulo,
Brazil} and Sadhan K Adhikari$^2$}

\address{$^1$Instituto de F\'{\i}sica, 
Universidade de S\~ao Paulo,\\
Caixa Postal 66318, 05315-970 S\~ao Paulo, Brazil\\
$^2$Instituto de F\'{\i}sica Te\'orica, 
Universidade Estadual Paulista,\\
01.405-900 S\~ao Paulo, S\~ao Paulo, Brazil\\}

\date{\today}
\maketitle

\begin{abstract}

We study the order parameter for mixed-symmetry states involving a major
$d_{x^2-y^2}$ state and various minor $s$-wave states ($s$, $s_{xy}$, and
$s_{x^2+y^2}$) for different filling and temperature for mixing angles 0
and $\pi/2$.  We employ a two-dimensional tight-binding model
incorporating second-neighbor hopping for tetragonal and orthorhombic
lattice. There is mixing for the symmetric $s$ state
both  on tetragonal and
orthorhombic lattice. The $s_{xy}$ state mixes with the $d_{x^2-y^2}$
state only on orthorhombic lattice.  The $s_{x^2+y^2}$ state never mixes
with the $d_{x^2-y^2}$ state. The temperature
dependence of the  order parameters is also studied.

{PACS number(s): 74.20.Fg, 74.62.-c, 74.25.Bt}

\end{abstract} 


\newpage

\section{Introduction}

After many theoretical and experimental investigations on the high-$T_c$
cuprates \cite{1}, with a high critical temperature $T_c$,   the symmetry
of their order parameter is not yet completely known. 
It is
now accepted \cite{2,3,4} that the cuprates are
quasi-two-dimensional
superconductors and at
higher temperatures 
the symmetry of the order parameter is of the
$d_{x^2-y^2}$ type.  However, many experiments \cite{5,5b,5c,5d,6,7,8,9}
and
related theoretical studies  \cite{11,13,jpc,a1,a2,13a,13b,14,rk,rk1}
suggest
that at lower temperatures the order parameter of the cuprates has a mixed
symmetry of the $d_{x^2-y^2}+\exp (i\theta)\chi $ type, where $\chi$
represents a minor component  with a distinct symmetry superposed on
the major component $d_{x^2-y^2}$.  From theoretical analysis the mixing
angle $\theta$ can have the values $0$, $\pi /2, \pi,$ and $ 3\pi/2$. For
mixing angles $\pi/2$ and $3\pi/2$ the time-reversal symmetry is broken.
  Four possible candidates for the minor
symmetry state $\chi$ are the $d_{xy}$, $s$, $s_{x^2+y^2}$ and $s_{xy}$
states.  From a group theoretical point of view these states belong to the
same irreducible representation of the orthorhombic point group.  However,
there is still controversy about the nature of the minor component and the
value of the mixing angle in different high-$T_c$ materials
\cite{11,13,jpc,a1,a2,13a,13b,14}.

The possibility of a mixed $(s-d)$ wave symmetry in superconductors was
suggested sometime ago by Ruckenstein et al. \cite{rk} and Koltiar
\cite{rk1}.
Several
phase-sensitive measurements on the order parameter indicate a significant
mixing of minor $s$-wave component with a major $d_{x^2-y^2}$ state at
lower temperatures in YBa$_2$Cu$_3$O$_7$ (YBCO).  In several
experimental analysis on the
Josephson supercurrent for tunneling between a conventional $s$-wave
superconductor (Pb) and twinned or untwinned single crystals of
YBCO, the possibility of mixed states like
$d_{x^2-y^2}\pm s$ or $d_{x^2-y^2}\pm is$ has been conjectured at lower
temperatures \cite{5,5b,5c,5d}.  More recently, Kouznetsov {\it et al.}
\cite{6}
performed $c$-axis Josephson tunneling experiments by depositing a
conventional superconductor (Pb)  across the single twin boundary of a
YBCO crystal and from a study of the critical current under an applied
magnetic field they also conjecture a similar mixed state in YBCO. From
the measurement of the microwave complex conductivity of high quality
YBa$_2$Cu$_3$O$_{7-\delta}$ single crystals at 10 GHz using a high-$Q$ Nb
cavity Sridhar {\it et al.} \cite{7} also suggested the possibility of
$d_{x^2-y^2}\pm s$ or $d_{x^2-y^2}\pm is$ states. In view of this
we perform a theoretical study of this problem using a tight-binding model
including orthorhombic distortion and second-nearest-neighbor hopping.

There is some consensous about $(d-s)$ mixing in YBCO, but in other
cuprates one could have a mixing between $d_{x^2-y^2}$ and $d_{xy}$
components. However,  we shall not study this mixing in this work. It is
pertinent to refer to the studies which led to this conclusion
in case of Bi$_2$Sr$_2$CaCu$_2$O$_{8}$ \cite{8,9}.

Although, there is no suitable microscopic theory for high-$T_c$ cuprates,
the existence of Cooper pair as the charge carrier is usually accepted
for these materials.
However, there continues controversy about a proper description of the
normal state and the pairing mechanism for such materials. More recently,
for underdoped systems it has not been possible to accomodate the
experimentally observed pseudogap  \cite{arp,arp1} above the
superconducting critical
temperature in a microscopic theory.  However, with the
increase
of doping
the superconducting critical temperature increases and the pseudo gap is
reduced and eventually it disappears at optimal doping.

It is generally accepted that in high-$T_c$ cuprates, superconductivity
resides mainly in the CuO$_2$ planes with nearly tetragonal symmetry.
Consequently, the critical temperature and superconducting gap of
high-$T_c$ materials are expected to be very sensitive to the level of
doping which determines the number of conduction electrons
in the two-dimensional CuO$_2$ plane of the cuprates and it is interesting
to study the effect of doping or filling on the superconductivity in
cuprates.  The present single-band model can acomodate a maximum filling
of two electrons (spin up and down) per unit cell. 
In addition, the experimental observation of 
a large anisotropy in the penetration depth between the $a$
and $b$ directions \cite{ba} in YBCO suggests an orthorhombic distortion
in YBCO and the present study is extended to include such a distortion.    

In the absence of
a microscopic theory, we use a phenomenological two-dimensional
tight-binding model with appropriate lattice symmetry for studying some of
the general features of the mixed-symmetry states involving $d_{x^2-y^2 
}$ and different $s$ states. This model has been
used successfully in describing many properties of high-$T_c$ materials
\cite{11,13,jpc,a1,a2,13a,13b}. 
Both orthorhombicity and filling are expected to play a crucial role in
the evolution of these mixed-symmetry states and we study in this paper
the properties of these  states  for different temperature,
filling,
orthorhombicity, and second-neighbor hopping.

In Sec. II we describe the formalism. In Sec. III we present our numerical
results and finally in Sec. IV we give a brief summary. 

\section{Theoretical Formulation}

The present tight-binding  model is sufficiently general for
considering 
mixed angular momentum states on tetragonal and orthorhombic lattice, 
employing nearest and second-nearest-neighbor hopping integrals.  
Here we take the  effective interaction $V_{\bf k
q}$ for transition from a momentum $\bf q$ to $\bf k$  to be
separable, and is expanded in terms of some general basis functions $\eta
_{i\bf k}$, labeled by index $i$, so that 
\begin{equation}\label{1}
V_{\bf k q}=- V_1 \eta _{1\bf k}\eta _{1\bf q}
- V_2 \eta _{2\bf k}\eta _{2\bf q}\end{equation}  
The separable nature of the interaction facilitates the solution of the
gap equation.
The orthogonal functions $\eta _{i\bf k}$ are associated with a 
one-dimensional irreducible representation of the point group of square
lattice $C_{4v}$ and are appropriate generalizations of the circular
harmonics incorporating the proper lattice symmetry. Here $V_i$ is the
coupling of effective interaction in the specific angular momentum state.
In the present investigation 
we  consider predominant 
singlet Cooper pairing and subsequent  condensation
in the $d$ and $s$ states, denoted by indices 1 and 2, respectively, and
the
mixed-symmetry state formed by these two. 

 The first function $\eta _{1\bf q}$
corresponding to the $d_{x^2-y^2}$ state is given by
\begin{eqnarray}
\eta _{1\bf q}& =& \cos q_x -\beta \cos q_y, \hskip 1cm  d_{x^2-y^2}
\mbox{-wave},
\end{eqnarray}
whereas the second function could be one of the following $s$ states
\begin{eqnarray}
\eta _{2\bf q}&=& 1,  \hskip 3.2cm s \mbox{-wave}, \label{c1}\\
    \eta _{2\bf q}&= &2\cos q_x \cos q_y,  \hskip 1.34cm  s_{xy}
\mbox{-wave}, \label{c2}\\
\eta _{2\bf q}&= &\cos q_x +\beta \cos q_y,  \hskip 1cm s_{x^2+y^2}
\mbox{-wave}, \label{c3}
\end{eqnarray} 
etc. 
Here $\beta  =1$ corresponds to square lattice and $\beta \ne 1$
represents orthorhombic distortion.  
The orthogonality property of functions $\eta$'s  is 
taken to be
\begin{eqnarray}\label{2}
\sum_{\bf q} \eta _{1\bf q}\eta _{2\bf q} = 0, ..., i\ne j.
\end{eqnarray} 
Property (\ref{2}) is approximate for choice (\ref{c3}) for 
$\beta\ne 1$.

We consider a single tight-binding two-dimensional band with an electron
dispersion relation including second-nearest-neighbor hopping. 
In this case the quasiparticle dispersion relation relating the electronic
energy $\epsilon_{\bf k}$  and momentum ${\bf k}$ is taken as 
\begin{equation}\label{3}
\epsilon_{\bf k}=-2t(\cos k_x+\beta \cos k_y-2\gamma\cos k_x
\cos k_y) -\mu,
\end{equation}
where $t$ and $\beta t$ are the nearest-neighbor hopping integrals
along the in-plane $a$ and $b$ axes, respectively, and $\gamma t$ is the 
second-nearest-neighbor hopping integral. In Eq. (\ref{3}) $\mu$ is the
chemical potential measured with respect to the Fermi energy and is
determined once the filling $n$ is specified.  The
nearest-neighbor hopping parameter $t$ is typically taken to be $\sim 0.1$
eV. The parameter $\beta$ destroys the symmetry between the $a$ and $b$
directions in the CuO$_2$ planes in this simple model. The potential
$V_{\bf k
q}$ above
also possesses such a symmetry-breaking term. 
The
energy $\epsilon_{\bf k}$ is measured with respect to the Fermi surface.
Such a one-band model with different first-neighbor-hopping parameters in
the $a$ and $b$ directions is the simplest approximate way of including in
the theoretical description the effect of orthorhombicity.

At a finite temperature $T$, one has  the following gap equation
\begin{eqnarray}
\Delta_{\bf k}& =& -\sum_{\bf q} V_{\bf kq}\frac{\Delta_{ \bf q}}{2E_{\bf
q}}\tanh
\frac{E_{\bf q} }{2k_BT}  \label{4} \end{eqnarray} 
with $E_{\bf q} =
[(\epsilon_{\bf q}-\mu  )^2 + |\Delta_{\bf q}|^{2}]^ {1/2},$ 
 and $k_B$ the Boltzmann constant. 

It has been observed that the critical temperature $T_c$ is sensitive to
the level of doping which determines the number of available conduction
electrons in the CuO$_2$ plane. In this model  
the chemical potential $\mu$ and the  the filling $n$ are 
determined by the number equation 
\begin{equation}\label{5}
n= 1- \sum_{\bf q}\frac {\epsilon_{\bf q}-\mu}{E_{\bf
q}}\tanh
\frac{E_{\bf q} }{2k_BT}.
\end{equation} 
The filling $n$ can be related to the the experimental doping $\delta$ in
the three dimensional Brillouin zone by $n=1-\delta$. In this work we
study the  variation of $n$ from 0 to 1 (half filling).

The order parameter $\Delta _{\bf q}$ has the
following anisotropic form: 
\begin{equation}\label{6}
\Delta _{\bf q} =  \Delta_1 \eta_{1\bf q} +C \Delta_2 \eta_{2\bf q},
\end{equation}
where $C\equiv \exp (i\theta) $ is a complex
number of unit modulus $|C|^2 = \cos^2\theta+\sin^2 \theta =1$.  
We substitute Eqs.  (\ref{1}) 
and (\ref{6}) into  the gap equation (\ref{4}) and using the orthogonality 
property (\ref{2}) obtain the two following coupled equations for
$\Delta_1$ and $\Delta_2$: 
\begin{eqnarray}\label{7}
\Delta_1 &=& \sum_{\bf q}V_1\eta_{1\bf q}\frac {\Delta_1 \eta_{1\bf q}
+C \Delta_2 \eta_{2\bf q}}{2E_{\bf q}}\tanh \frac{E_{\bf q}}{2k_BT},\\
C\Delta_2 &=& \sum_{\bf q}V_2\eta_{2\bf q}\frac {\Delta_1 \eta_{1\bf q}
+C \Delta_2 \eta_{2\bf q}}{2E_{\bf q}}\tanh \frac{E_{\bf q}}{2k_BT}.
\label{8}
\end{eqnarray}

Equations (\ref{7}) and (\ref{8}) can be substantially simplified for a
purely imaginary $C$, e.g., for $C=\pm i$ ($\theta =\pm \pi/2$). In this
case for real $\Delta_1$ and $\Delta_2$, the real and imaginary parts of 
these equations become, respectively,
\begin{eqnarray}\label{10}
\Delta_1 &=& \sum_{\bf q}V_1\frac {\Delta_1 \eta^2_{1\bf q}
}{2E_{\bf q}}\tanh \frac{E_{\bf q}}{2k_BT},\\
\Delta_2 &=& \sum_{\bf q}V_2\frac {
 \Delta_2 \eta^2_{2\bf q}}{2E_{\bf q}}\tanh \frac{E_{\bf q}}{2k_BT},
\label{11}
\end{eqnarray}
since in this case
\begin{equation}
E_{\bf q} =
[(\epsilon_{\bf q}-\mu  )^2 +  \Delta_1^2 \eta^2_{1\bf q} + \Delta_2^2
\eta^2_{2\bf q}]^ {1/2},\label{9}
\end{equation} 
\begin{eqnarray}\label{12}
 \sum_{\bf q} \frac { \eta_{1\bf q}\eta_{2\bf q}
}{2E_{\bf q}}\tanh \frac{E_{\bf q}}{2k_BT}=0,
\end{eqnarray}
which follows from the definitions of $\eta_{1\bf q}$ and $\eta_{2\bf q}$
and Eq.  (\ref{9}). However, Eq. (\ref{12}), which is responsible for the
simplification for $C=\pm i$ ($\theta= \pi/2$ and $3\pi/2$), does not hold
for $C=\pm 1$ or for a general
complex $C$. 

For $C=\pm 1$ ($\theta= 0,\pi$)  no further simplification of the
coupled Eqs. (\ref{7}) and (\ref{8}) is possible and one has 
\begin{eqnarray}\label{7a}
\Delta_1 &=& \sum_{\bf q}V_1\eta_{1\bf q}\frac {\Delta_1 \eta_{1\bf q}
\pm \Delta_2 \eta_{2\bf q}}{2E_{\bf q}}\tanh \frac{E_{\bf q}}{2k_BT},\\
\pm\Delta_2 &=& \sum_{\bf q}V_2\eta_{2\bf q}\frac {\Delta_1 \eta_{1\bf q}
\pm \Delta_2 \eta_{2\bf q}}{2E_{\bf q}}\tanh \frac{E_{\bf q}}{2k_BT},
\label{8a}
\end{eqnarray}
with 
\begin{equation}\label{7aa}
E_{\bf q} =
[(\epsilon_{\bf q}-\mu  )^2 + (\Delta_1 \eta_{1\bf q} \pm \Delta_2
\eta_{2\bf q})^2]^ {1/2},
\end{equation}

Finally, for a general complex $C$ one can separate Eqs.  (\ref{7}) and
(\ref{8}) into their real and imaginary parts. In this case Eq. (\ref{12})
is not valid, and the above procedure results in four equations for the
two
unknowns $\Delta_1$ and $\Delta_2$.  These four equations
are consistent only if $\Delta_1 =0$ or $\Delta_2 =0$, which means that
there could not be mixing between the two components. So
mixed-symmetry states are allowed only for mixing angles $\theta = 0,
\pi/2, \pi, $ and $3\pi/2$, or for $C = \pm 1, $ and  $\pm i$ and we shall
consider only these cases in the following.

The ultraviolet momentum-space divergence of the Bardeen-Cooper-Schrieffer
 equation was
originally neutralized by a physically-motivated Debye cut off \cite{e}. 
This procedure had the advantage of reproducing the experimentally
observed isotope
effect. 
It can also be handled by using the technique of renormalization
\cite{5a1,5a2}.  
Here we introduce a cut off in the momentum sums of the gap
equation. As there is no
pronounced isotope effect in the high-$T_c$ cuprates, the present cut off
is merely a mathematical one without any reference to the phonon-induced
Debye cut off.
   In Eqs. (\ref{7})  and (\ref{8})
both the interactions $V_1$ and
$V_2$ are assumed to be energy-independent constants for $|\epsilon_{\bf
q} - \mu| < k_B T_D$ and zero for $|\epsilon_{\bf q} - \mu| > k_B T_D$,
where $k_B T_D$ is the present  cut off.

\section{Numerical Result}

We solve the coupled set of equations (\ref{10}) and (\ref{11})  or
(\ref{7a})  and (\ref{8a})  in conjunction with the number equation
(\ref{5}) numerically and calculate the gaps $\Delta_1$ and $\Delta_2$ at
various filling and temperature.  This gives us the opportunity to study
the mixed-symmetry states for different filling and temperature on both
tetragonal and orthorhombic lattice.   Throughout the present study  we
consider the cut off $k_B T_D=0.1t $  with the parameter $t=0.2586$ eV.
This corresponds to a cut-off of $T_D = 300$ K. The order parameters 
$\Delta_{x^2-y^2}$, $\Delta_{xy}$ and  $\Delta$  presented in this work
are all in
units of $t$.

We study the mixture of the $d_{x^2-y^2}$ and the symmetric $s$
state (= 1)  on tetragonal and orthorhombic lattice with
second-nearest-neighbor hopping contribution. The results of our study
have interesting variation as the second nearest hopping parameter is
varied and this is studied in detail in the following for mixing angles 0
and $\pi/2$.  First we consider the coupled $d_{x^2-y^2}+is$ wave at $T=0$
on a tetragonal lattice. In
this case Eqs. (\ref{10}) and (\ref{11}) are applicable. The parameters
for this model on tetragonal lattice are the following: $\beta=1$,
$V_{1}=0.73t$, and $V_2=1.8t$. In Fig. 1(a) we plot
$\Delta_1\equiv\Delta_{x^2-y^2}$ and $\Delta_2\equiv\Delta_{s}$ 
for different filling $n$ and for $\gamma =0$, 0.05, 0.1, and 0.2.
The mixing between $d_{x^2-y^2}$ and $s$ states takes place for 
values of $n$ close to half filling.  For $\gamma =0.2$ the $d$ wave is
completely suppressed
and we have pure $s$ wave order parameter for all $n$.  For orthorhombic
distortion the parameters of the model are $\beta=0.95$, $V_{1}=0.97t$,
and
$V_2=2.1t$.  In Fig. 1(b) we plot $\Delta_{x^2-y^2}$ and $\Delta_{s}$
on orthorhombic lattice for
different filling $n$ and for $\gamma =0$, 0.05, 0.1, and 0.2. The
qualitative nature of the order parameters in Figs. 1 (a) and (b)  are the
same although there are quantitative differences.  In both cases the
mixing  is  limited to large values of $n$. However, in
Fig. 1(b) there is mixing for $\gamma = 0.2$, whereas there is none in
Fig. 1(a). In Fig. 1(a) we observe a gradual decrease in $d$-wave 
order parameter at half filling with increase in $\gamma$. The mixing
depends very much on doping and second nearest neighbour hopping.

We study superconductivity in coupled $d_{x^2-y^2}+s$ wave at $T=0$,
governed by Eqs. (\ref{7a}) and (\ref{8a}), which corresponds to the
mixing angle $0$. On a tetragonal lattice there is no mixing but   
meaningful mixing is possible on
orthorhombic lattice and we study this case in detail.
In Fig. 2 we plot $\Delta_{x^2-y^2}$ and
$\Delta_{s}$  for different filling $n$ and
$\gamma =0$, 0.05, 0.1, and 0.2 calculated with $\beta=0.95$,
$V_{1}=0.97t$, and $V_2=2.1t$.  In this case the $d$ and the $s$ waves can
coexist and the mixing occurs for large $n$ values.  However, with
increasing $\gamma$ the $d$-wave component is reduced in magnitude.

Next we consider the mixing of the $d_{x^2-y^2}$ wave with the $s_{xy}$
wave for both mixing angles 0 and $\pi/2$. In both cases there is no
mixing on tetragonal lattice. However, there is mixing on a orthorhombic
lattice and we discuss the detail below. First we consider the
$d_{x^2-y^2}+is_{xy}$ case, where we solve Eqs. (\ref{10}) and (\ref{11}) 
with $\beta= 0.95$, $V_{1}=0.95t$, and $V_2=1.17t$, for $\gamma=0,
0.05,0.1$ and 0.2. In Fig. 3(a) we plot $\Delta_{x^2-y^2}$
and $\Delta_{s_{xy}}$  for different filling
$n$.  The interesting region of mixing occurs for very large values of
$n$.  Next we consider the $d_{x^2-y^2}+s_{xy}$ case, where we solve Eqs.
(\ref{7a}) and (\ref{8a})  with $\beta= 0.95$, $V_{1}=0.95t$, and
$V_2=1.17t$, for $\gamma=0, 0.05,0.1$ and 0.2. In Fig. 3(b) we plot
$\Delta_{x^2-y^2}$ and $\Delta_{s_{xy}}$ 
 for different filling $n$.  No mixing is found between
$\Delta_{x^2-y^2}$ and $s_{x^2+y^2}$ waves on tetragonal and orthorhombic
lattice and we shall not discuss this case further.

Now we investigate the temperature dependence of the order parameters in
different cases. We studied several cases for different values of $n$,
$\gamma$ and $\beta ( = 1, 0.95)$. The qualitative nature of the
temperature dependence in different cases are different and we discuss
them separately. Figure 4(a)  illustrates the temperature dependencies of
the
order parameters for the $d_{x^2-y^2}+is$ case on square lattice
calculated with the parameters of Fig. 1(a) for $\gamma=0.05$ and
$n=0.95$
(critical temperature 69 K)  and for $\gamma=0.1$ and $n=0.9$ (critical
temperature 70 K). The nature of the order parameters of Fig. 4(a) does
not
change in the presence of orthorhombic distortion. The order parameters in
this case are similar to those in the uncoupled case. The only difference
is that at lower temperatures in the presence of the $s$-wave component
the $d$-wave order parameter gets a bit suppressed. In this case the
$s$-wave order parameter goes to zero at a temperature lower than $T_c$.

In Fig. 4(b) we show the temperature dependencies of the order
parameters for the $d_{x^2-y^2}+s$ case on orthorhombic lattice calculated
with the parameters of Fig. 2 for $\gamma=0.05$ and $n=0.9$ (critical
temperature 70 K),
 and for $\gamma=0$ and $n=0.95$ (critical temperature 90 K). There is a
qualitative difference between the order parameters of Figs. 4(a) and (b). 
In Fig. 4(b) both the components become zero at the critical temperature
and a mixed-wave order parameter is present for all temperatures below
$T_c$, whereas in Fig. 4(a) there is a temperature region where only the
$d$-wave order parameter exists.

Next we study the order parameters in case of mixture with the $s_{xy}$
state. First we consider the $d_{x^2-y^2}+s_{xy}$ type mixture on
orthorhombic lattice corresponding to Fig.  3(b) with $\gamma =0.1$,
$n=0.95$,
$T_c=98$ K, and with $\gamma =0.2$, $n=0.95$, $T_c=64$ K. The order
parameters for different temperatures are plotted in Fig. 5(a).  
The temperature dependence in this case behaves as in Fig. 4 (b). Both
components are nonzero immediately  below the critical temperature. Next
we consider the $d_{x^2-y^2}+is_{xy}$ type mixture
on orthorhombic lattice corresponding to Fig.  3(a) with $\gamma =0.1$,
$n=0.9$, $T_c=68$ K, and with $\gamma =0.0$, $n=0.98$, $T_c=73$ K. The
corresponding order parameters at different temperatures are plotted in
Fig. 5(b).  The nature of the order parameters in Fig. 5(a) is quite
different
from those in Fig. 5(b).
In all cases we observe very different temperature dependencies of order 
parameter compared to the standard BCS-model results for uncoupled wave.

\section{Summary}

In this work we have studied  the mixed-symmetry
superconducting states comprising of 
$d_{x^2-y^2}$ and different $s$ waves appropriate for 
two-dimensional cuprates using a tight-binding model on tetragonal and
orthorhombic lattice.
We studied the variation of order parameters
with filling $n$ for tetragonal and
orthorhombic lattices for different second-neighbor hopping for pure and 
mixed-symmetry states. The mixing of $d_{x^2-y^2}$ and $s$ waves varies 
considerably with doping and second-neighbor hopping. We observe mixing
to
take place at large values of filling.
 We have also
studied the temperature dependence of the order parameter
under different situations. The temperature dependence for
$d_{x^2-y^2}+is$ is similar to our previous studies \cite{a1,13a} for
the same mixing at $n=1.$ However,  this
dependence is very different for the time reversal symmetry
cases $d_{x^2-y^2}+s$ and $d_{x^2-y^2}+s_{xy}$.

We thank Conselho
Nacional de Desenvolvimento Cient\'{\i}fico e Tecnol\'ogico and Funda\c c\~ao
de Amparo \`a Pesquisa do Estado de S\~ao Paulo for financial support.


{\bf Figure Captions:}
\vskip 1cm

1. The order parameters $\Delta_{x^2-y^2}$ (dashed line) and $\Delta_s$
(full line) for mixed wave $d_{x^2-y^2}+is$ on (a) tetragonal and (b) 
orthorhombic lattice for different $n$ and $\gamma$. The parameters for
the model on tetragonal lattice are $\beta =1$, $V_1=0.73t$, $V_2=1.8t$
and on orthorhombic lattice are $\beta = 0.95$, $V_1=0.97t$, $V_2=2.1t$.

2. The order parameters $\Delta_{x^2-y^2}$ (dashed line) and $\Delta_s$
(full line) for mixed wave $d_{x^2-y^2}+s$ on  
orthorhombic lattice for different $n$ and $\gamma$. The parameters for
the model  are $\beta = 0.95$, $V_1=0.97t$, $V_2=2.1t$.

3. The order parameters $\Delta_{x^2-y^2}$ (dashed line) and
$\Delta_{s_{xy}}$
(full line) for mixed wave (a) $d_{x^2-y^2}+is_{xy}$ and (b) 
 $d_{x^2-y^2}+s_{xy}$ on  
orthorhombic lattice for different $n$ and $\gamma$. The parameters for
both models (a) and (b)  are $\beta = 0.95$, $V_1=0.95t$, $V_2=1.17t$.

4. Temperature dependence of the order parameters $\Delta_{x^2-y^2}$
and $\Delta_{s}$ for mixed-symmetry 
(a) $d_{x^2-y^2}+is$  state on
tetragonal
lattice for  $\gamma =0.05$  and $n=0.95$ (dashed line) and 
for  $\gamma =0.1$  and $n=0.9$ (full line)  and 
for mixed symmetry 
(b) $d_{x^2-y^2}+s$  state on
orthorhombic 
lattice for  $\gamma =0.05$  and $n=0.9$ (dashed line) and 
for  $\gamma =0$  and $n=0.95$ (full line). 

5. Temperature dependence of the order parameters $\Delta_{x^2-y^2}$
and $\Delta_{s}$ for mixed-symmetry 
(a) $d_{x^2-y^2}+s_{xy}$  state on
orthorhombic 
lattice for  $\gamma =0.1$  and $n=0.95$ (dashed line) and 
for  $\gamma =0.2$  and $n=0.95$ (full line)  and 
for mixed symmetry 
(b) $d_{x^2-y^2}+is_{xy}$  state on
orthorhombic 
lattice for  $\gamma =0.1$  and $n=0.9$ (dashed line) and 
for  $\gamma =0$  and $n=0.98$ (full line).

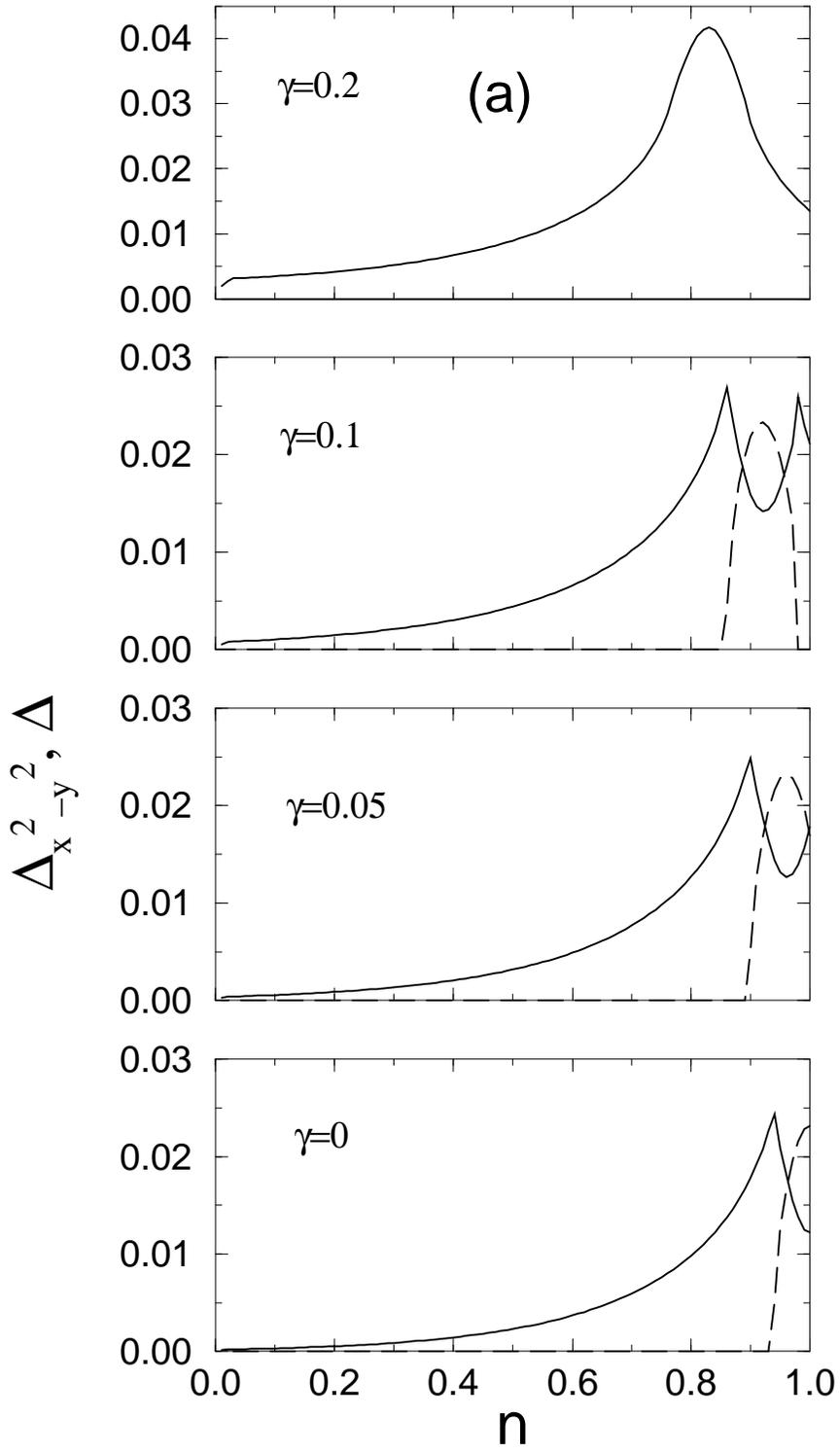
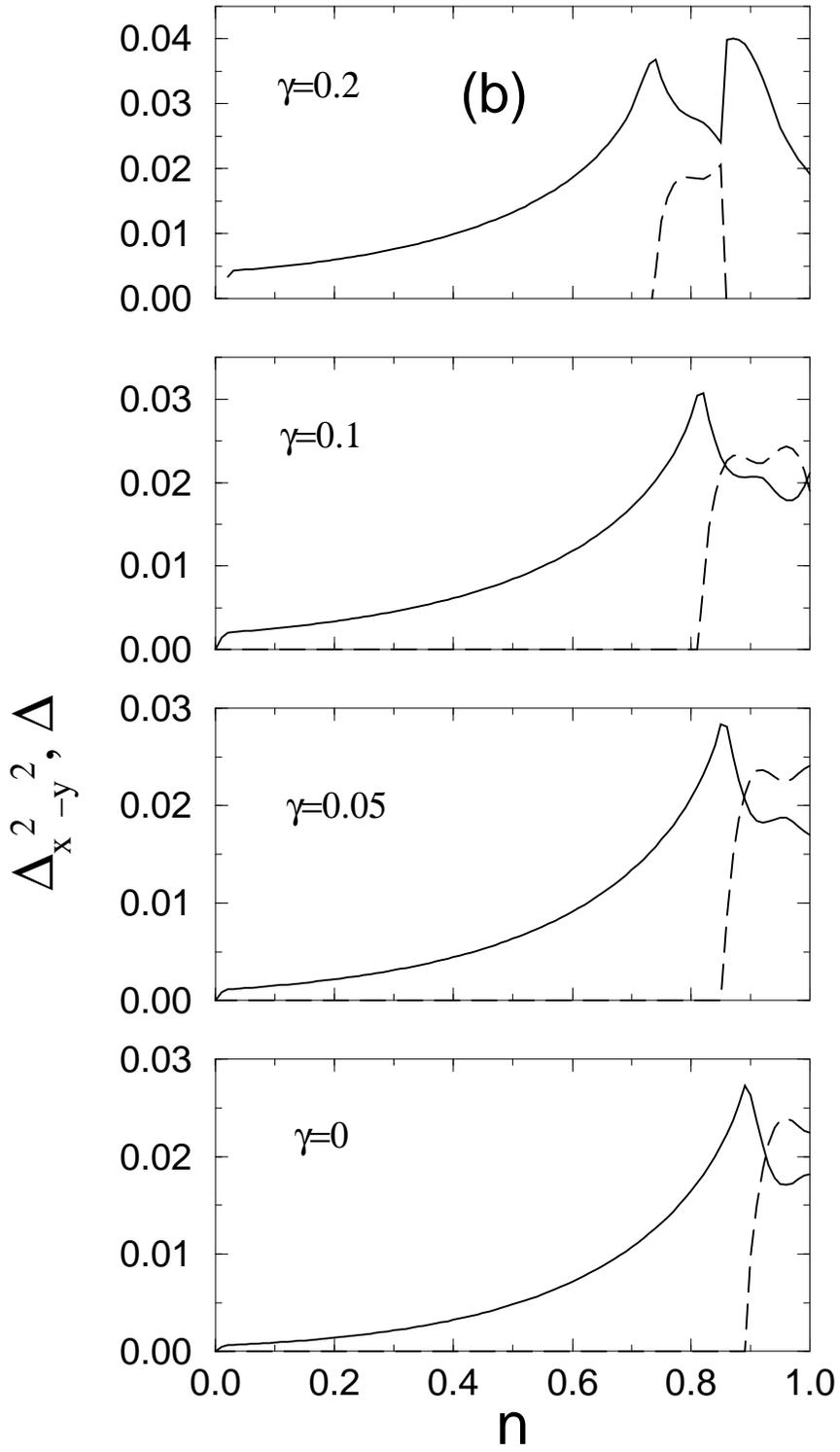
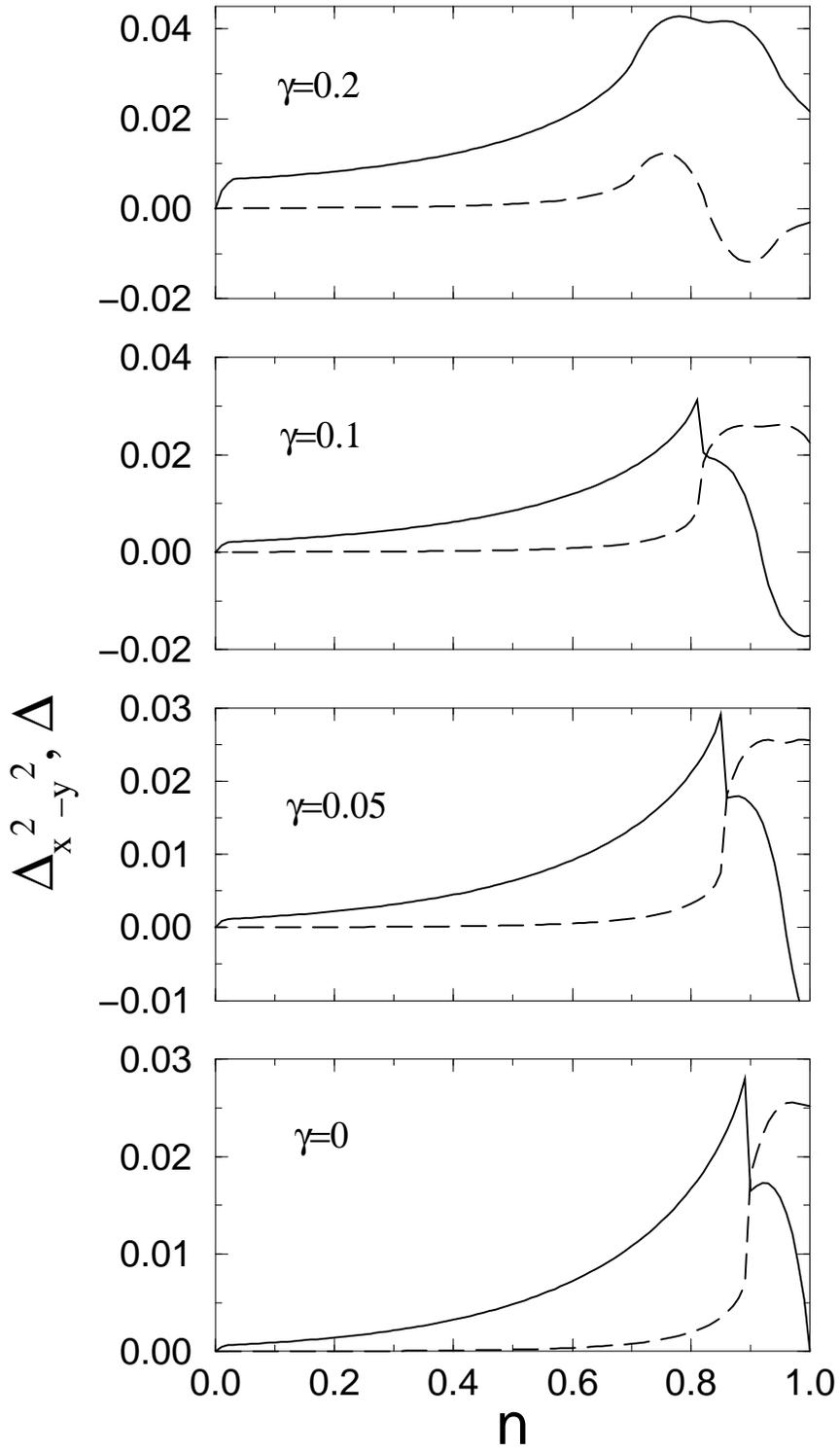
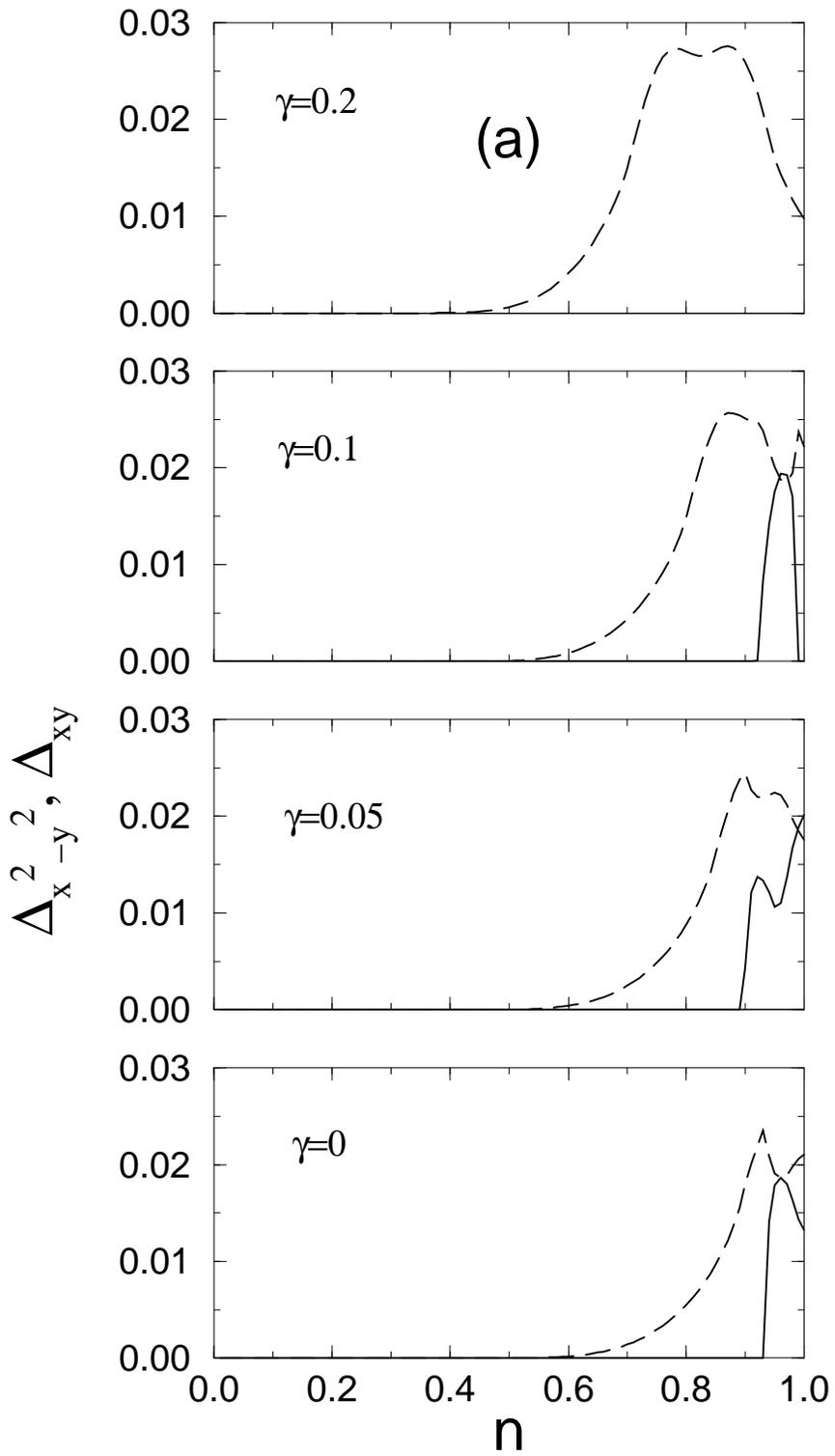
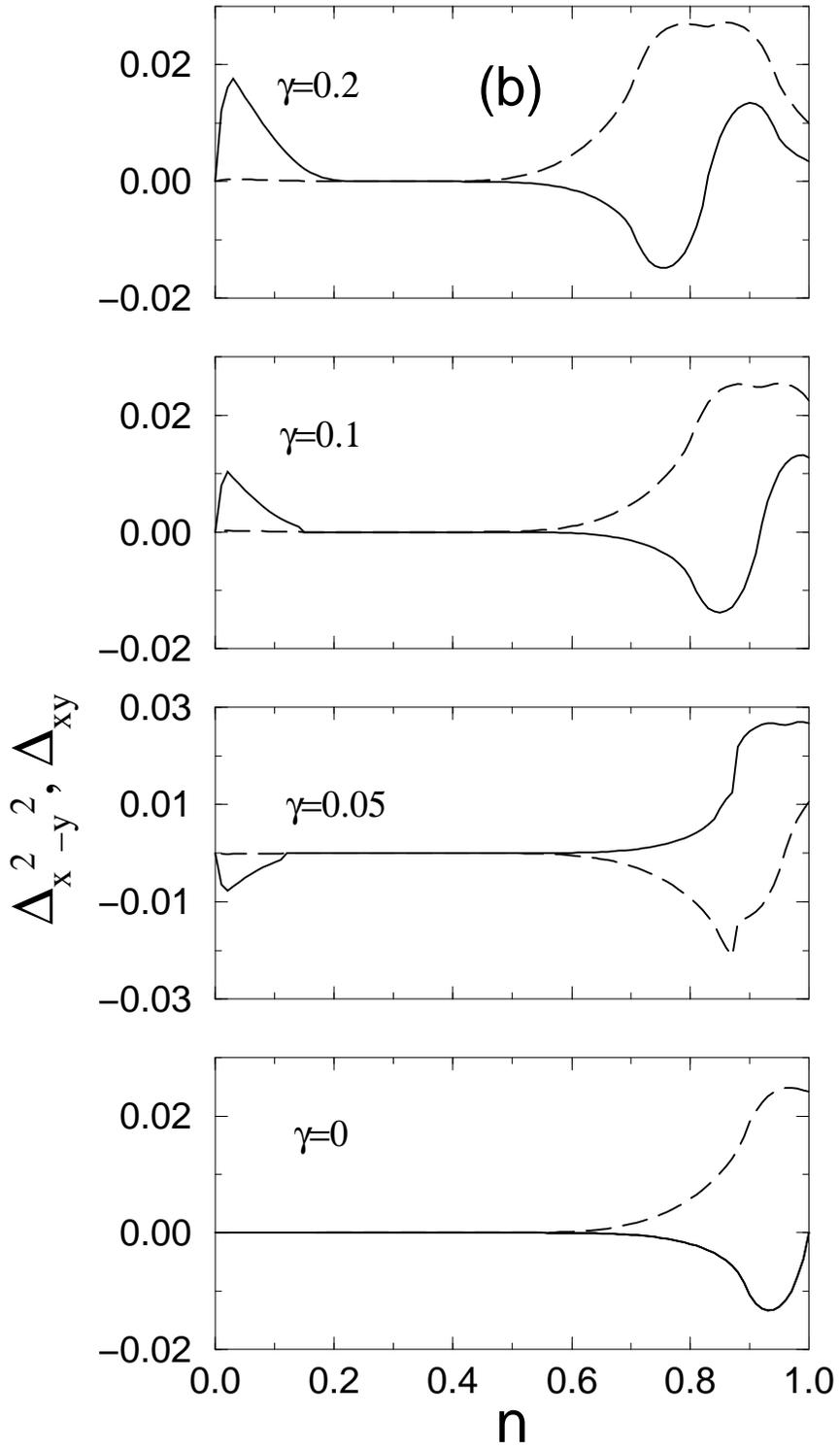
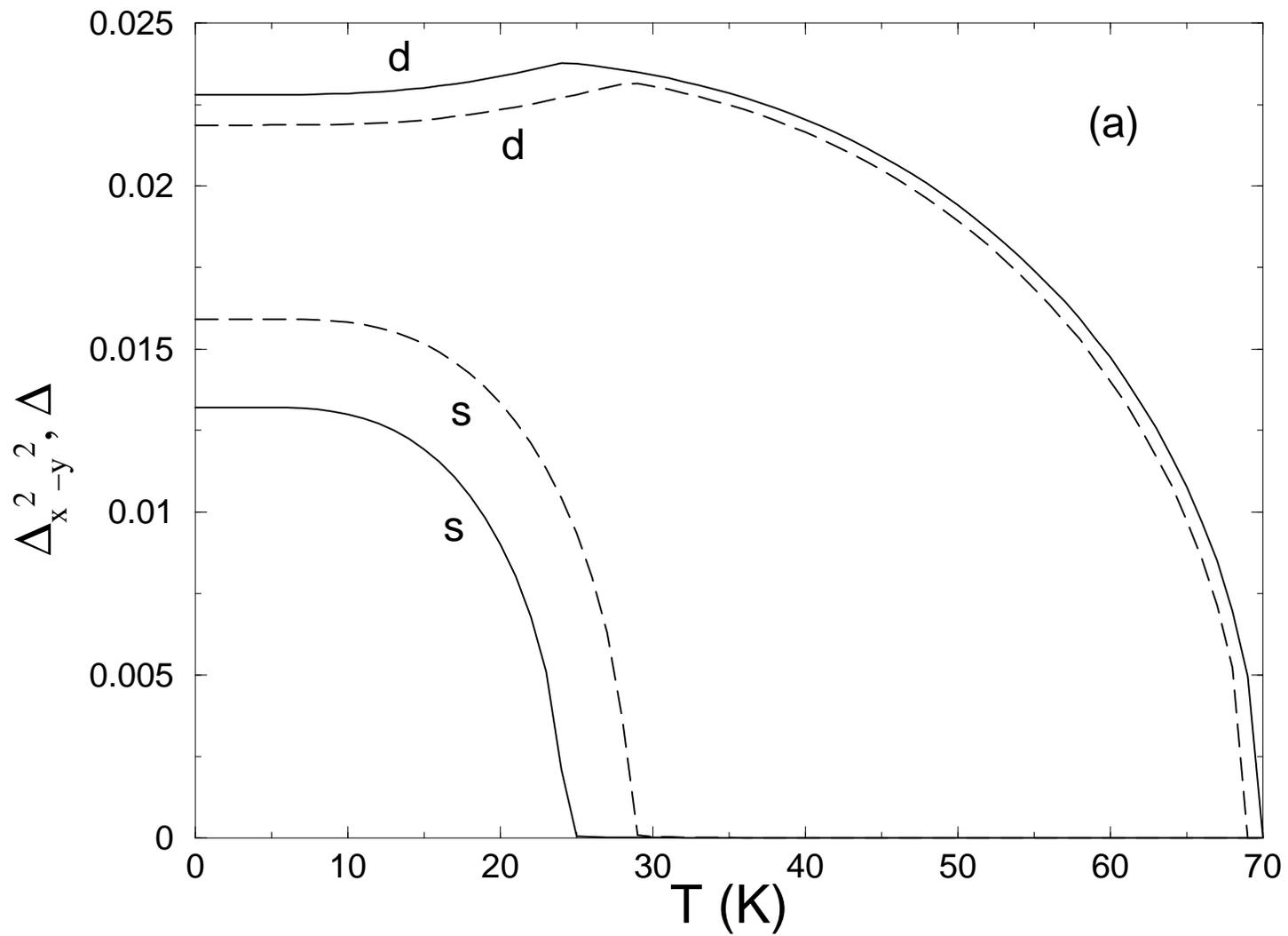
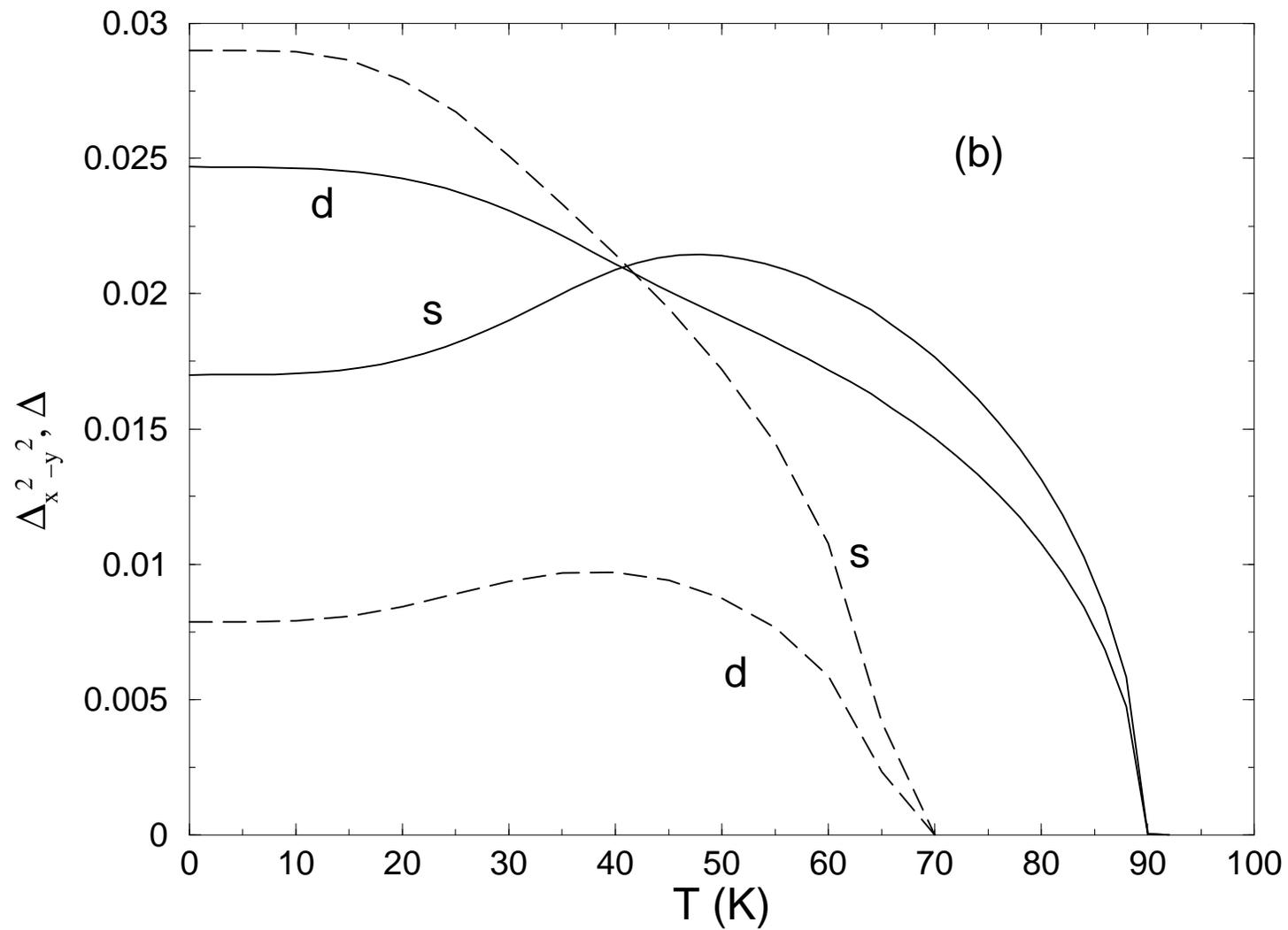
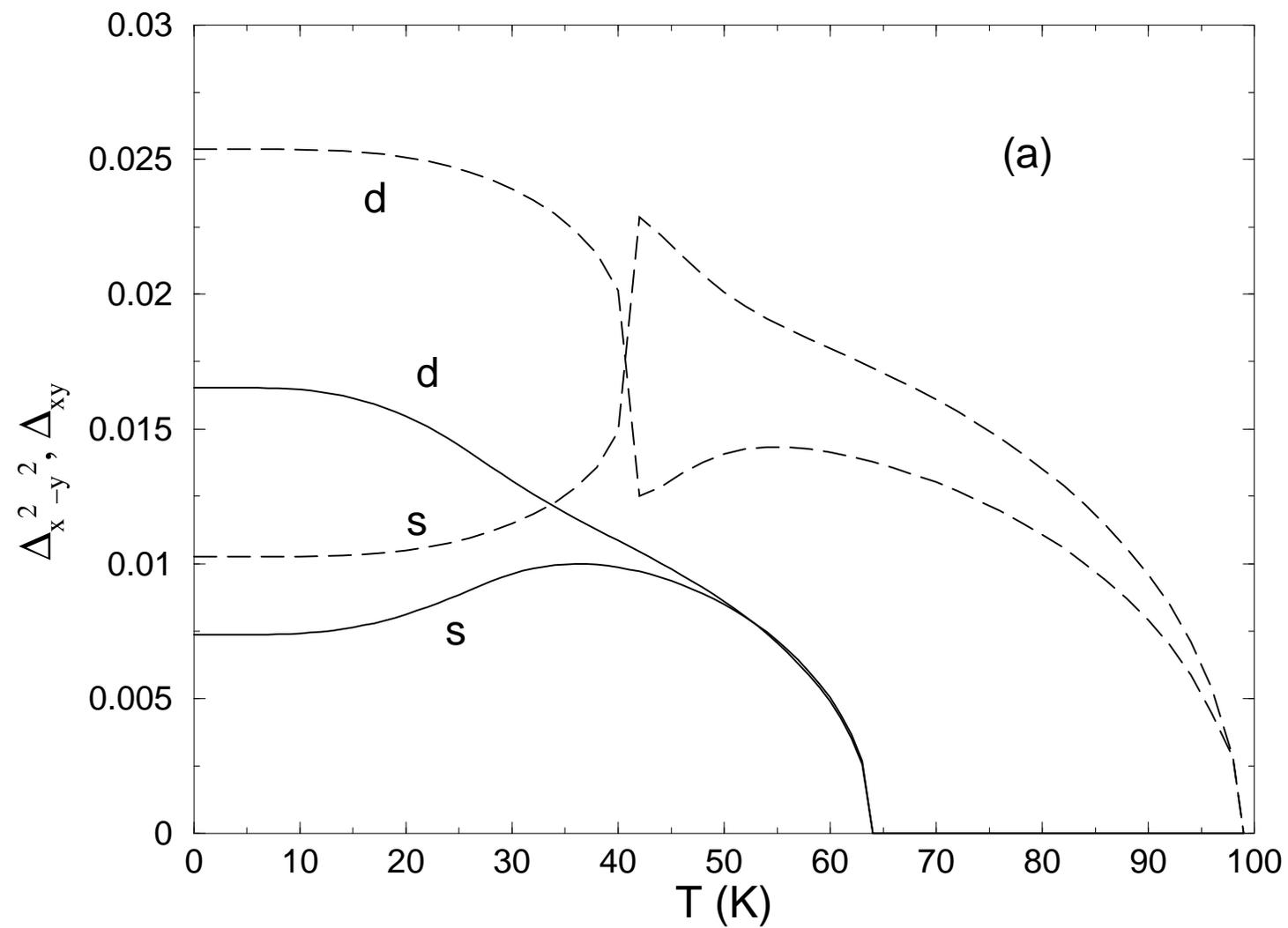
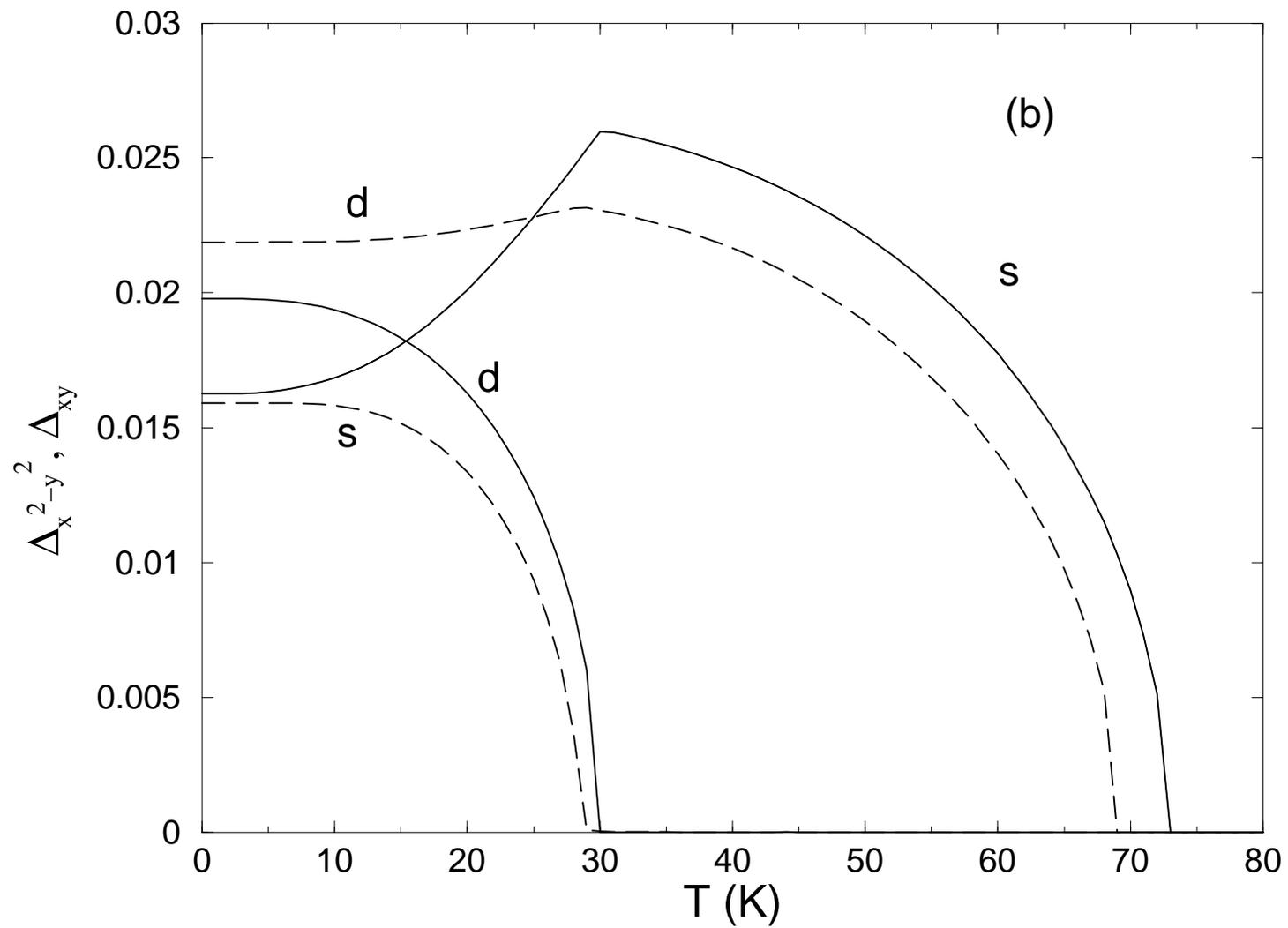
\end{document}